\documentclass[useAMS]{mn2e}
\usepackage{graphicx}

\newbox\grsign \setbox\grsign=\hbox{$>$} \newdimen\grdimen
\grdimen=\ht\grsign
\newbox\simlessbox \newbox\simgreatbox
\setbox\simgreatbox=\hbox{\raise.5ex\hbox{$>$}\llap
     {\lower.5ex\hbox{$\sim$}}}\ht1=\grdimen\dp1=0pt
\setbox\simlessbox=\hbox{\raise.5ex\hbox{$<$}\llap
     {\lower.5ex\hbox{$\sim$}}}\ht2=\grdimen\dp2=0pt

\newbox\simppropto
\setbox\simppropto=\hbox{\raise.5ex\hbox{$\sim$}\llap
     {\lower.5ex\hbox{$\propto$}}}\ht2=\grdimen\dp2=0pt

\title
  {The nature of extremely red galaxies in the local universe}
\author[L. Sodr\'e Jr., A. Ribeiro da Silva, W. A. Santos]
  {Laerte Sodr\'e Jr.\thanks{laerte@astro.iag.usp.br}$^1$,
   Aline Ribeiro da Silva$^1$ and Walter A. Santos$^2$ \\
  $^1$Departamento de Astronomia, IAG/USP, Rua do Mat\~ao 1226, 05508-090, 
 Cidade
  Universit\'aria, S\~ao Paulo, SP, Brazil \\
  $^2$Department of Physics and Astronomy, University College London, 
Gower Street, London WC1E 6BT, UK}

\begin{document}

\maketitle

\begin{abstract}
We investigate the nature of extremely red galaxies  (ERGs), objects whose 
colours are redder than those found in the red sequence present in 
colour-magnitude diagrams of galaxies. We selected from the Sloan 
Digital Sky Survey Data Release 7 a volume-limited sample of such 
galaxies in the 
redshift interval $0.010 < z < 0.030$, brighter than $M_r=-17.8$ 
(magnitudes dereddened, corrected for the Milky Way extinction) and 
with $(g-r)$ colours larger than those of galaxies in the red sequence. 
This sample contains 416  ERGs, which were 
classified visually. Our classification was cross-checked
with other classifications available in the literature. We found from
our visual classification that the majority of objects in our sample 
are edge-on spirals ($73$\%). Other spirals correspond to $13$\%, 
whereas elliptical galaxies comprise only $11$\% of the objects. 
After comparing the morphological mix and the distributions of 
H$\alpha$/H$\beta$ and axial ratios of ERGs and objects in the
red sequence, we suggest that dust, more than stellar 
population effects, is the driver of the red colours found in 
these extremely red galaxies.

\end{abstract}

\begin{keywords}
 galaxies: fundamental parameters - galaxies: photometry -
galaxies: spiral - surveys: galaxies
\end{keywords}

\section{Introduction}
The bimodality is a conspicuous feature of optical colour-magnitude diagrams 
(CMDs) of galaxies in the local and distant universe (e.g., Strateva et al. 
2001; Kauffmann et al. 2003; Wiegert, de Mello \& Horellou 2004; Mateus et al. 
2006; Nicol et al. 2011), and is able to provide
interesting constraints on how galaxies evolve (Mateus 2007; Asari et al. 2007;
Taylor et. al. 2011). It presents two major features: 
the red sequence, populated mostly by passive, "dead", galaxies, 
in general ellipticals, lenticulars and passive spirals, and the blue cloud, 
containing spirals and irregulars with
ongoing star-formation activity. The region between these two major features
is often called the green valley (e.g., Baldry et al. 2004; Mendez et al. 
2011; Gon\c calves et al. 2012). The colour bimodality has an
environmental component: while the blue star-forming galaxies tend to populate
low-density regions, red-sequence galaxies are often found in clusters and
rich groups (e.g., Oemler 1974; Dressler 1980; Postman \& Geller 1984). 

The position of a galaxy in a CMD is often interpreted in terms of its
evolutionary status and, in particular, on whether or not, or at which level,
it is still forming new stars. Indeed, spectrophotometric models suggest 
that, after $\sim$1 to 2 Gyr, a blue galaxy which had its star formation 
stopped by internal or environmental mechanisms migrates from the blue 
cloud to the red sequence (e.g., Bell et al. 2004; Blanton 2006; Gabor et al. 
2010), crossing the green valley. While most of these galaxies would have 
already changed their morphology from late-type to early-type galaxies, a 
significant number of spirals can also be found in the red sequence 
(e.g., Bamford et al. 2009, Skibba et al. 2009, Masters et al. 2010b, 
Robaina et al. 2012, Tojeiro et al. 2013), 
especially at high redshifts (Bundy et al. 2010, Bell et al. 2012).
These results suggest that colour is much more sensitive to environment
than morphology, with colour transformations from blue to red occurring on
time-scales much shorter than those of morphological transformations.

But a close inspection of optical CMDs of galaxies in the local universe show
many galaxies with colours above those of the red sequence. Why are these
galaxies so red? Is this due to old, probably metal rich stars? Is this 
due to dust? Indeed, dust in the interstellar medium of galaxies, besides 
absorbing part of the optical light, may redden galaxy colours. There is,
actually, an age-extinction degeneracy: 
a galaxy possessing high extinction can have
colors similar to an older object without extinction (Worthey 1994, de 
Meulenaer et al. 2013). Dust affects mostly star forming objects (e.g., Alam \&
Ryden 2002, Masters et al. 2010a), because star formation occurs in
dusty molecular clouds in discs.
The effects of dust on the observed properties of spirals, including 
colours, have been studied as a function on their inclination 
(e.g., Cunow 1992; Giovanelli et al. 1994; Tully et al. 1998; Masters, 
Giovanelli \& Haynes 2003; Masters et al. 2010a), and, as expected,  
are more significant for edge-on spirals.

The objective of this paper is to address the nature of the extremely
red galaxies (ERGs)- those above the ordinary red sequence of passive galaxies-
in the local universe. We want to know what galaxies are these, their 
morphology,  and why their colours are so red, if due mainly 
to stellar populations or to dust. We approach this
problem by selecting a sample of nearby ERGs from the Sloan Digital Sky Survey
(SDSS) Data Release 7 (Abazajian et al. 2009) and examining 
visually their images.
We also compare our visual classification with those produced by the Galaxy
Zoo project (Lintott et al. 2011) and by the automated support vector
machine algorithm of Huertas-Company et al. (2011). 

It is worth mentioning that our ERGs are not the same as the extremely
red objects (EROs), found in optical and in infrared surveys at larger
redshifts and often associated with galaxies dominated by old populations 
or dusty starbursts (e.g., Cimatti et al. 2004; Kong et al. 2009) although
some ERGs, if at high redshift, could be classified as EROs.

The outline of this paper is as follows. We present, in Section 2,
the SDSS sample of ERGs compiled in this work. Section 3 describes the
results of our visual classification and a comparison with classification
obtained by other authors. In Section 4 we discuss the origin of the extreme 
colours of these galaxies. Finally, in Section 5, we summarize our 
main findings.

\section{The data}
To investigate the nature of the ERGs in the nearby universe, we have
selected objects classified as galaxies and with measured spectroscopic redshift from the Sloan Digital
Sky Survey (SDSS) Data Release 7 (Abazajian et al. 2009), in the redshift
interval $0.010 < z < 0.030$ and brighter than $r=17.77$ (the spectroscopic 
magnitude limit). 
This selection criterion was adopted for two reasons:
first, the low redshift is required 
for having galaxies close enough and with relatively large apparent size to 
assure a good resolution for visual inspection of the images,
minimizing morphological misclassifications; second, the redshift interval
should be narrow enough to avoid $k-$corrections and evolutionary effects
affecting the galaxy colours.

We also imposed the following photometric quality flags in the galaxy
selection: nchild = 0, not BRIGHT, not SATURATED,
not SATUR\_CENTER, in order to minimize the number of objects with image 
defects returned by the query.

This selection leads to 31067 galaxies. 
The colour-magnitude diagram (CMD), $(g-r)$ vs $M_r$, of the selected galaxies 
is shown in Figure 1. All apparent magnitudes here are of the model type, 
which should give more reliable colour 
measurements\footnote{http://www.sdss.org/dr7/algorithms/photometry.html}
than other types of magnitudes, and are corrected by Galaxy extinction 
(dereddened magnitudes). 
Absolute magnitudes were computed assuming a $\Lambda$CDM
cosmological model with $H_0 = 72$ km s$^{-1}$ Mpc$^{-1}$, $\Omega_m=0.3$ and
$\Omega_\Lambda=0.7$. We did not apply any $k$-correction in the estimation of
absolute magnitudes because, in the redshift interval considered here, they are
very small and even smaller than their own uncertainties (Blanton \&
Roweis 2007). The CMD shown in Figure 1 presents a conspicuous
red-sequence ($g-r \sim 0.8$), as well as the blue cloud ($g-r \sim 0.4$).

We consider here  as extremely red galaxies (ERGs) those above the line
\[ (g-r) = 0.81-0.03(M_r+18.1067) \]
in the CMD; this is the line above the red sequence in Figure 1.
We adopt here a volume limited sample, by considering only galaxies with
luminosities above $M_r=-17.76$. There are 468 objects above the line in our
volume-limited sample. 

It is important to verify whether the extreme colours found here are real
or outliers/artifacts in SDSS. Indeed, an examination of the
$(g-r) \times (r-i)$ colour-colour diagram for this sample reveals that
$\sim 10$\% of the objects are possibly outliers. They were removed by 
adopting colour cuts
as follows: $(g-r) < 1.2$ and $0.3 < (r-i) < 1.0$. The colour-colour
diagram obtained after removing 52 outliers is shown in Figure 2. It reveals 
that most galaxies follow a sequence in this diagram. The total number of 
ERGs in the sample is 416. Notice that galaxies in this sample
are fairly distributed in SDSS area, not presenting any relevant bias with
respect to Galaxy extinction.

The results of the next sections are relatively
robust with respect to the definition of extremely red galaxies. For example,
considering a sample of galaxies in Figure 1 with $(g-r) > 0.9$, we obtain
qualitatively the same results described in the next sections.

\begin{figure}
\includegraphics[scale=0.4]{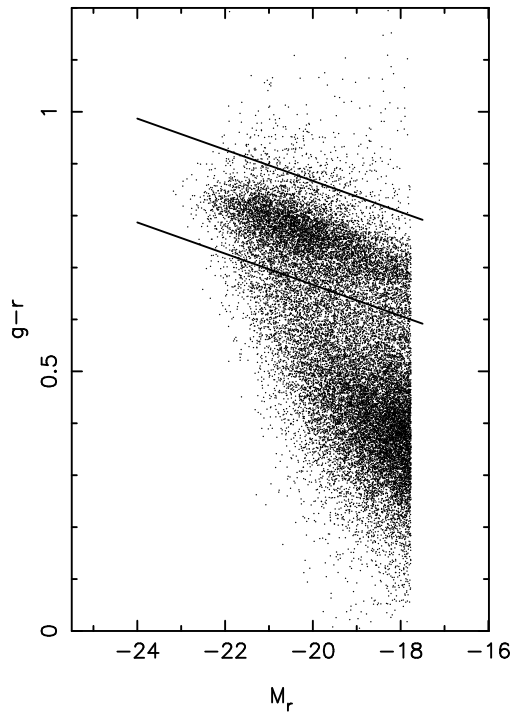} 
\caption{Colour-magnitude diagram ($g-r$ vs $M_r$) for our galaxy sample.
The galaxies above the top straight line
are defined here as the ERGs. We consider as members of the red sequence the 
galaxies between the two straight lines.}
\label{cmd}
\end{figure}

\begin{figure}
\includegraphics[scale=0.4]{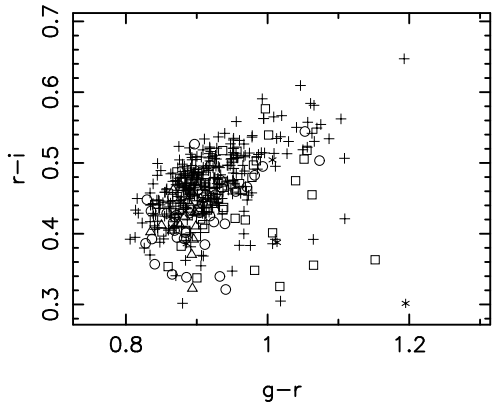} 
\caption{Colour-colour diagram ($g-r$ vs $r-i$) for our ERG sample.
The diagram includes all galaxies above the line adopted to define
extremely red galaxies which survived our colour-cut, 
$g-r < 1.2$ and $0.3 < r-i < 1.0$. The different symbols correspond to the 
classes in Table \ref{table_visual}: 1- crosses, 2- squares, 3- circles, 
4- triangles, 5- asterisk.}
\label{cc}
\end{figure}

\section{Morphological analysis}
To investigate the morphology of the galaxies in our ERG sample we
adopted three procedures: a) direct visual inspection of images; b) comparison
with Galaxy Zoo classification (Lintott et al. 2011), 
and c) comparison with Huertas-Company
et al. (2011) automated classification. The results are presented in the next 
subsections.

\subsection{Classification by visual inspection}
All ERG images were examined with the SDSS task 
Navigator\footnote{http://cas.sdss.org/dr7/en/tools/chart/navi.asp}, 
which provides composite colour images of SDSS objects. Each image was
then classified in one of the following classes:
1) spiral galaxy seen edge-on, 2) spiral galaxy with visible 
spiral arms or noticeable inclination, 3) elliptical or lenticular galaxy, 
4) merger/interacting/irregular object, 
and 5) image with defects, with bad segmentation or with
Milky Way stars projected on to the central parts of their image,  affecting 
the galaxy colours. Example images for those 5 classes are given
in Figure \ref{hahb}.

Despite the low redshift of this
sample, the classification of more compact galaxies is somewhat uncertain;
for example, it is sometimes difficult to 
distinguish a face-on compact spiral blurred by the seeing from an elliptical
or lenticular galaxy. Edge-on spirals and lenticulars are also difficult to
distinguish, although sometimes evidence for star formation can be seen in
the image. A face-on lenticular and an elliptical are also
hard to distinguish; although our class 3 is dominated by elliptical galaxies,
it probably also contains some lenticulars. 
Merger here comprises a large class of objects with
irregular morphology due mostly to interactions and mergers. The ``defects''
(class 5) include all those images which, for some reason, are not useful 
for our analysis.

The results of our classification are summarized in Table \ref{table_visual}. 
By far, most of the ERGs in our sample are edge-on objects ($73\pm4$\%).
The fractions of (not edge-on) spiral and elliptical plus lenticular galaxies
are roughly similar, $\sim 12$\%, about six time smaller than that of edge-on 
discs. We have assumed Poissonian errors here and throughout this paper.

\begin{table}
\begin{center}
 \caption{Results of our visual classification of the ERG sample.}
 \label{table_visual}
 \begin{tabular}{cccc}
  \hline
class & type & number & fraction (\%) \\
  \hline
1 & Edge-on spirals & 304 & 73$\pm$4 \\
2 & Spirals with visible arms &  52 &  13$\pm$2 \\
3 & Ellipticals and Lenticulars   & 44 & 11$\pm$2 \\
4 & Merger/Interacting  & 12 & 3$\pm$1 \\
5 & Defects & 4 & 1$\pm$1 \\
  \hline
 \end{tabular}
\end{center}
\end{table}

\begin{figure}
\includegraphics[scale=0.5]{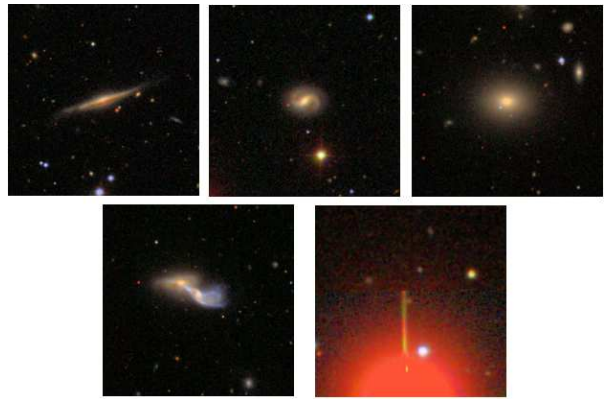} 
\caption{Examples of the 5 morphological classes adopted in the visual 
classification in this work.
Top, from left to right: spiral edge-on, spiral with visible arms, elliptical
and/or lenticular.
Bottom, from left to right: merger/interacting, image defect.}
\label{class}
\end{figure}

\subsection{Classification by Galaxy Zoo}
We cross-checked our classification with that provided by the
Galaxy Zoo project and discussed in Lintott et al. (2011). This ``citizen 
science" project provides morphological classifications (actually 
probabilities that a galaxy is in a given morphological class) of nearly
900 thousand galaxies produced by hundreds of thousands volunteers. This
classification has proven useful in many analysis requiring morphological 
information (e.g., Bamford et al. 2009, Skibba et al. 2009, Darg et
al. 2009, Schawinski et al. 2009, Masters et al. 2011, Hoyle et al. 2012). 

There are many potential biases that can affect the morphological 
classifications produced by Galaxy Zoo, which were discussed by Bamford et al. 
(2009) and  Lintott et al. (2011). These biases occur because visual 
classifications are affected by the 
brightness and apparent size of the galaxies in the sample: naturally, smaller
and fainter objects are more difficult to classify than larger and brighter
objects. As a consequence, for each object the Galaxy Zoo team provides
two types of classification. The first one comprises the classes
 elliptical (E), edge-on spirals (S edge-on), spirals with visible arms
(S other; a combination of Galaxy Zoo classes clockwise spirals, CW, and 
anticlockwise spirals, ACW),  Merger and Don't Know, with the 
fraction of volunteers votes for each class (notice that
the "edge-on" category from Galaxy Zoo includes both proper edge-on spirals 
and also those without clear arms). 
The second type comprises debiased votes, where the classification bias 
mentioned above is statistically corrected for
(see Appendix A of Bamford et al. 2009 for details). The debiased classes are
designated as E$_d$ for ellipticals and S$_d$ for spirals (CW, ACW and 
edge-on).

We adopted the maximum probability to ascribe a morphological type to a galaxy.
Out of 416 objects in our sample, there are 409 and 408 with ordinary and
debiased classifications in Galaxy Zoo, respectively.  
This classification, summarized in Table \ref{galaxy_zoo},
also demonstrates that edge-on objects are dominant among ERGs, 
comprising 65$\pm$4\% of the sample. Considering the debiased votes, the
table indicates that about 1/4 of the ERGs were classified as ellipticals
whereas 3/4 were classified as spirals.

\begin{table}
\begin{center}
 \caption{Galaxy Zoo classification}
 \label{galaxy_zoo}
 \begin{tabular}{ccccc}
  \hline
class & number & fraction (\%) & fraction (\%) \\
      &  ERG   &   ERG & RS & Notes\\
  \hline
S edge-on & 267 & 65$\pm$4 & 23$\pm$1 & 1\\
S other & 21 & 5$\pm$2 & 14$\pm$1 & 1\\
E & 100 & 24$\pm$2 & 60$\pm$1 & 1 \\
Merger & 13 & 3$\pm$1 & 1$\pm$1 & 1 \\
Don't Know & 8 & 2$\pm$1 & 1$\pm$1 & 1 \\
      &     &   & & \\
E$_d$ & 101 & 25$\pm$2 & 56$\pm$1 & 2 \\
S$_d$ & 307 & 75$\pm$4 & 44$\pm$1 & 2 \\
  \hline
 \end{tabular}
Notes: 1- ``raw'' votes; 2- debiased votes.
\end{center}
\end{table}

\subsection{Classification by SVM}
Huertas-Company et al. (2011) performed an automated classification of
about 700,000 galaxies from the DR7 spectroscopic sample using a support
vector machine (SVM) algorithm, estimating the probability of a galaxy
being in each of four morphological types: E, S0, Sab and Scd. All but one
galaxies in our sample have a SVM classification.

Ascribing again the most
probable type to a galaxy, we obtain the fractions summarized in Table
\ref{svm}. We verify that a bit more than 3/4 of the ERG sample (79$\pm$4\%) 
is constituted by
spirals, with the remaining fraction containing mostly lenticular galaxies
(16$\pm$1\%) and ellipticals (4$\pm$1\%). Unfortunately, in this classification 
there is not a distinction between edge-on and spirals with visible arms.

The SVM algorithm found $16$\% of lenticular galaxies in our sample.
There is a notorious difficulty for distinguishing lenticular galaxies 
from ellipticals or spirals in images
like those of SDSS. In our visual examination, non edge-on galaxies without
evidence of spiral arms were included in class 3. We verified that, 
considering only galaxies
classified as S0 by the SVM method, half of them were classified by us as
edge-on spirals and the other half was split more or less equally between
ellipticals and spirals with visible arms.

\begin{table}
\begin{center}
 \caption{SVM classification.}
 \label{svm}
 \begin{tabular}{cccc}
  \hline
class & number & fraction (\%) & fraction (\%) \\
      &  ERG   &   ERG & RS \\
  \hline
E & 17 & 4$\pm$1 & 21$\pm$1  \\
S0 & 67 & 16$\pm$1 & 28$\pm$1 \\
Sab & 331  & 79$\pm$4 & 50$\pm$1 \\
Scd & 0 & 0 & 0 \\
  \hline
 \end{tabular}
\end{center}
\end{table}

\section{The driver of the extreme red colours}
The results of the Galaxy Zoo and SVM classifications presented in the
previous section indicate that most of the galaxies in our ERG sample are 
spirals, in agreement with our visual classification.
Actually, our results and the Galaxy Zoo classification show that the
majority of our ERG sample is comprised by edge-on galaxies.

Between 2/3 and 3/4 of the galaxies above the red sequence are discs seen 
edge-on, as 
demonstrated by our visual classification (73\%) and by the Galaxy Zoo 
direct voting (65\%). What makes these galaxies so red? Effects of dust
or of stellar populations? Indeed, galaxies may become redder either by 
increasing their extinction by dust or due to the presence of an intrinsically
very red, old and/or high-metallicity population (see, e.g., the illustration of
Bruzual \& Charlot 2003 SSP spectra in Figure 1 of Cid Fernandes et al. 2005). 
This behaviour actually reflects the existence of an 
extinction-age-metallicity degeneracy.

To shed light on the nature of the extreme colours of ERGs, it is useful
to know the morphological composition of galaxies pertaining to the red
sequence (RS). 

For this exercise, we consider as members of the RS those
galaxies between the two
straight lines shown in Figure \ref{cmd}. 
The upper line was used in Section 2 for
defining the ERGs. The lower line is 0.2 bluer in $(g-r)$. There are 
8635 objects of our sample in the RS, most of them with Galaxy Zoo
and SVM classifications. The morphological mix of the RS sample is
also shown in Tables \ref{galaxy_zoo} and \ref{svm}, and is 
quite different from the ERG mix. For the RS, the classical early-type
galaxies- E and S0- dominate, the opposite of what is found for the ERG sample,
which is dominated by spirals. But many spirals are also found in the RS, 
$\sim 40-50$\%. Many of these red spirals might be examples of passive 
spirals as described by Masters et al. (2010b), but the red sequence also
includes edge-on and dust-reddened spirals and, probably, spirals with large, 
red bulges. There are no late-type spirals (Scd) in the
ERG and RS samples: they are in the blue cloud.

The clear differences in the morphological mix of the ERG and RS samples may
be interpreted as an evidence that extinction by dust, and not stellar 
population effects, is the main driver of the extreme colours of the 
objects in the ERG sample. The results in Tables \ref{galaxy_zoo} and 
\ref{svm} are consistent with the concept of the RS being populated by 'dead 
galaxies', those without significant star formation at least over the 
last 1-2 Gyr, irrespective of their morphological types. 
If the ERG extreme colours were due to intrinsically redder stars (due to 
extreme age or metallicity), we should expect that the dominant population
would be comprised by ordinary early-type galaxies, not spirals as observed 
in our sample, since there is not any evidence that passive spiral galaxies 
harbour stellar populations which are intrinsically redder than those in
ellipticals due to age or metallicity effects. 
Masters et al. (2010a) show that the 
typical $(g-r)$ reddening from face-on to edge-on spirals is about 0.15 
magnitudes. So, for a spiral galaxy be in the ERG region of the CMD, it
probably started in the red sequence, being a passive spiral. However, given 
the large variance in the reddening difference between face-on and edge-on 
spirals (see Figure 9 of  Masters et al. 2010a), it
is not impossible that even star-forming spirals may be members of the ERG 
sample if seen edge-on (see Tojeiro et al. 2013 for a discussion on on-going
star formation in red galaxies). Anyway, using the NED extinction calculator,
one can verify that the variation in the extinction in $(g-r)$ for the Milky 
Way is larger than unit for Galactic latitudes lower than $\sim$5 degrees 
from the Galactic plane.

\begin{figure}
\includegraphics[scale=0.5]{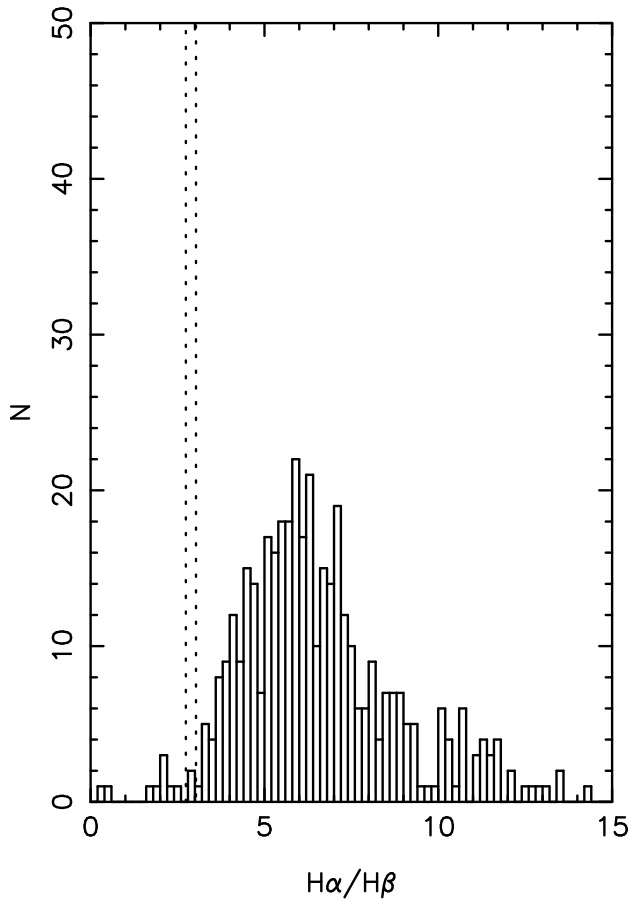} 
\caption{Distribution of the emission line intensity ratio H$\alpha$/H$\beta$
for galaxies in the ERG sample. The dotted lines correspond to the interval
expected for galaxies without intrinsic reddening.}
\label{hahb}
\end{figure}

We have also verified whether the nebular extinction, as measured by the
flux ratio H$\alpha$/H$\beta$, is large in our ERG sample. The intrinsic (extinction-free)
value of this ratio is insensitive to the physical conditions of the gas
where the line emission is produced, ranging from 3.03 for a gas temperature of
5000 K to 2.74 at 20000 K (Osterbrock 1989). We have used here the values
of this ratio as measured by the {\em STARLIGHT} code (Cid Fernandes et al.
2005), which takes in to account the intrinsic line absorption (modelled
with Bruzual \& Charlot 2003 spectra). The distribution of  H$\alpha$/H$\beta$
for the 379 galaxies in the ERG sample with this ratio measured is presented
in Figure \ref{hahb}, and is a clear indication that large extinctions are 
indeed present in this sample. Notice that some objects (mostly early-types) 
may have very low line emission and, consequently, large errors in this line 
ratio. The median value for H$\alpha$/H$\beta$ is 6.2, corresponding
to $A_V=2.1$ (assuming the Cardelli, Clayton \& Mattis 1989 extinction 
law)\footnote{In this
case, $A_V \simeq 6.31 \log[(H\alpha/H\beta)/2.86]$ (Stasi\'nska et al. 2004).}.
For comparison, considering only red sequence galaxies, we obtain a median 
value for H$\alpha$/H$\beta$ of 4.6, corresponding to $A_V=1.3$. This exercise
is also consistent with dust being the major driver of the colours of ERGs.

\begin{figure}
\includegraphics[scale=0.5]{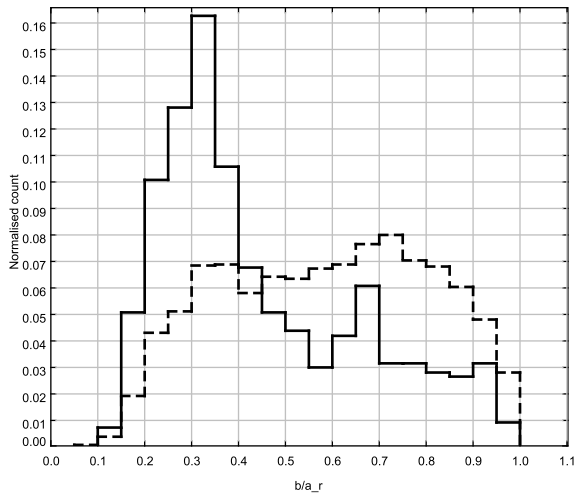} 
\caption{ Distribution of the model-type apparent
axis ratio $b/a$ of galaxy images in 
the $r$-band (see text for details). Continuum line: ERG sample; dashed line: 
RS sample.}
\label{ba}
\end{figure}

Besides the differences in the morphological mix of the ERG and RS samples,
there is also a clear difference in the fraction of edge-on objects: they are
almost 3 times more frequent among ERGs than in the RS.
To examine further this point, we present in Figure
\ref{ba} the distribution of the model minor to major axis ratio 
$b/a$ of galaxy images in 
the $r$-band for the ERG and RS samples. We adopt here ratios based 
on photometric fitting models, either exponential or de 
Vaucouleurs, ascribing to each galaxy the model with larger likelihood
(see Alam \& Ryden 2002 and the SDSS Algorithms 
Page\footnote{www.sdss.org/dr7/algorithms/} for details), although the form
of the distributions do not depend strongly of the model choice.
The $b/a$ distributions in Figure \ref{ba} show that the ERG sample has a 
significantly large number of galaxies with low values of axial ratios (around 
$b/a \sim 0.30$) in comparison with the RS distribution, 
which presents a mostly flat 
distribution. These results are in good agreement with what was found
by Alam \& Ryden (2002) and are consistent with the predominance of edge-on 
galaxies in the ERG sample and less-flat, bulge-dominated galaxies in the RS 
sample ($b/a \sim 0.75$).

The prevalence of edge-on spirals (or low values for the axial ratios) in our
sample of extreme objects seems a clear evidence that geometry is playing a 
significant role, as expected if the reddening is due to extinction by dust 
in galaxy disks. 

It is also worth mentioning that most not-edge-on spirals in the 
ERG sample also 
present prominent dust lanes. Some can be real passive spirals 
(Masters et al. 2010b), but not all, since some of them which seems strongly 
reddened in their centres, present blue disks indicative of ongoing star 
formation. Some ellipticals in our ERG sample also show significant dust lanes.

\section{Summary}
In this work we have selected a sample of extremely red galaxies, ERGs, with
colours redder than those found in the red sequence present in the 
colour-magnitude diagram of galaxies. We have verified that
most of the ERGs are edge-on spirals and that there are many more edge-on 
spirals in the ERG sample than among red sequence galaxies.

We propose that the reddest galaxies in our local universe 
have their extreme colours due mainly to the presence of dust. Many spiral
galaxies seen edge-on in our sample may be passive and have an intrinsically red
population, and extinction by dust places them above the red sequence. 
Additionally, even star-forming spirals seen edge-on may suffer enough 
extinction to be seen above the red sequence. On the other hand, it is unlikely
that the extreme red colours of the ERG sample is produced by
exceptionally old or high-metallicity red stars. 

Reddening by dust seems to be 
a natural consequence of the prevalence of edge-on spirals in our sample
of extremely red objects.

\section*{Acknowledgments}
The authors are grateful to an anonymous referee whose comments led to a
significant improvement in the presentation of our results. We also thank
William Schoenel for his help with the {\em STARLIGHT} data used in this work.
We thank the Brazilian agencies FAPESP and CNPq for supporting this work, which
was part of the undergraduate project of ARS. The authors 
also wish to thank the team of the SDSS for their dedication to a
project which has made the present work possible.

Funding for the Sloan Digital Sky Survey (SDSS) and SDSS-II has been provided by the Alfred P. Sloan Foundation, the Participating Institutions, the National Science Foundation, the U.S. Department of Energy, the National Aeronautics and Space Administration, the Japanese Monbukagakusho, and the Max Planck Society, and the Higher Education Funding Council for England. The SDSS Web site is http://www.sdss.org/.

The SDSS is managed by the Astrophysical Research Consortium (ARC) for the Participating Institutions. The Participating Institutions are the American Museum of Natural History, Astrophysical Institute Potsdam, University of Basel, University of Cambridge, Case Western Reserve University, The University of Chicago, Drexel University, Fermilab, the Institute for Advanced Study, the Japan Participation Group, The Johns Hopkins University, the Joint Institute for Nuclear Astrophysics, the Kavli Institute for Particle Astrophysics and Cosmology, the Korean Scientist Group, the Chinese Academy of Sciences (LAMOST), Los Alamos National Laboratory, the Max-Planck-Institute for Astronomy (MPIA), the Max-Planck-Institute for Astrophysics (MPA), New Mexico State University, Ohio State University, University of Pittsburgh, University of Portsmouth, Princeton University, the United States Naval Observatory, and the University of Washington.

\label{lastpage}

\end{document}